**Cooperative Problem Solving: an experience of high-school teaching updating**

Carlo Maria Carbonaro[1], Alessia Zurru[1,3], Viviana Fanti[1,2], Matteo Tuveri[2], Gianluca Usai[1,2]

[1]Department of Physics, University of Cagliari, Monserrato (Italy)

[2]INFN Sezione di Cagliari, Monserrato (Italy)

[3]Laboratorio Scienza, Department of Physics, University of Cagliari, Monserrato (Italy)

We present the results of an experience of teaching updating dispensed to Italian high school physics teachers to promote the application of the Cooperative Problem Solving method as an useful strategy to improve physics learning at high school level and to foster the development of problem solving skills. Beside analysing the method and discussing the ways to propose and apply it in a high school context, the teachers experienced the method acting both as learners and as tutors of student group learners. Students and teachers evaluated as positive the experience, mainly focusing on cooperation within the group by information exchange and the application of a solution scheme. The ex-post analysis of the students' performance in applying the method to solve some rich context text showed the need of improving critical sense with respect to achieved results to fully exploit the strategy and develop their problem solving skills. Finally, an analysis on gender differences and scholar distribution of students is presented.

**Keywords**: cooperative problem solving, gender analysis, high-school physics teachers, teaching physics

**Introduction**

As many other countries all around the world experienced[1,2] in the last decades Italy faced reduced enrolment in STEM (Science, Technology, Engineering and Mathematics) studies and, mainly, in hard natural sciences such as physics among the others. Since 2004 the Italian Ministry of University and Research promoted a national project (PLS, Piano Lauree Scientifiche[3]) aimed to increase the number of high school students pursuing enrolments and graduation in physics by means of a series of actions devoted to both students and teachers. Within this context, we promoted a course on Cooperative Problem Solving (CPS) to update knowledge of physics teachers and promote the diffusion of the method to teach physics in the high-school. It is indeed well known that problem solving is a skill strongly requested in the whole STEM courses and increasingly appreciated in professional and social world[4-8], being recognized as a habitus useful to manage new situations and contexts. Problem solving can be in general defined as the ability of one person to cope with a problem, the latter being a new situation which requires elaborating previous knowledge and experience to achieve the solution[9, 10]. Among scientists, physicists always valued problem solving as one of the most peculiar features of their discipline and spent a lot of efforts to analyse how to teach it and how to use it for teaching physics[11-17]. Teaching problem solving strategies to students was demonstrated as very effective in improving their performances in problem solving and their ability, in general, to use structured strategies to deal with professional issues[18-20]. Among the numerous methods experimented to teach problem solving[8, 17] Heller proposed to implement the Polya's solving strategy[21] in cooperative grouping[12, 13], focusing on cooperation as a key feature in the learning process. Cooperative learning was indeed proven successful at high school and college level in improving students' achievements and teaching approach[22-24]. In the present work we present the results of an experience of teaching updating dispensed to Italian high school physics teachers to promote the application of the Cooperative Problem Solving (CPS) method as a useful strategy to improve physics learning at high school level and to foster the development of problem solving skills. We also experimented with a group of teachers and one of students the approach to evaluate their willingness towards the method. Finally, we show the results of two different analysis about gender differences and scholar distribution of students and their performances in the cooperative problem approach.

**Methodology**

To promote the application of CPS approach in high school we organized a short course for teachers where the approach was explained and tested. Beside receiving the instruction of the approach (12 hours), the teachers discussed motivations and ways to apply CPS in a high school class and participated to few hours of exercitation where the teachers itself acted as members of the group (6 hours). Finally, the approach was tested with two mixed large classes of high school students, the teachers being involved as active coaches or as passive scouts (4 hours).

There were 40 teachers attending the course, 86% of them coming from scientific high school, 10% from technical schools, the rest from classical ones. The 98 students participating to the CPS laboratories were mainly IV and V years students (but ten from third year) and came largely from scientific high school (84%, the rest from classical high school).

To analyse the experience, we questioned the appeal of the approach and its applicability as foreseen by the teachers and we asked the students to evaluate their experience in solving some text-enriched problems. Finally, we analysed the results of the students.

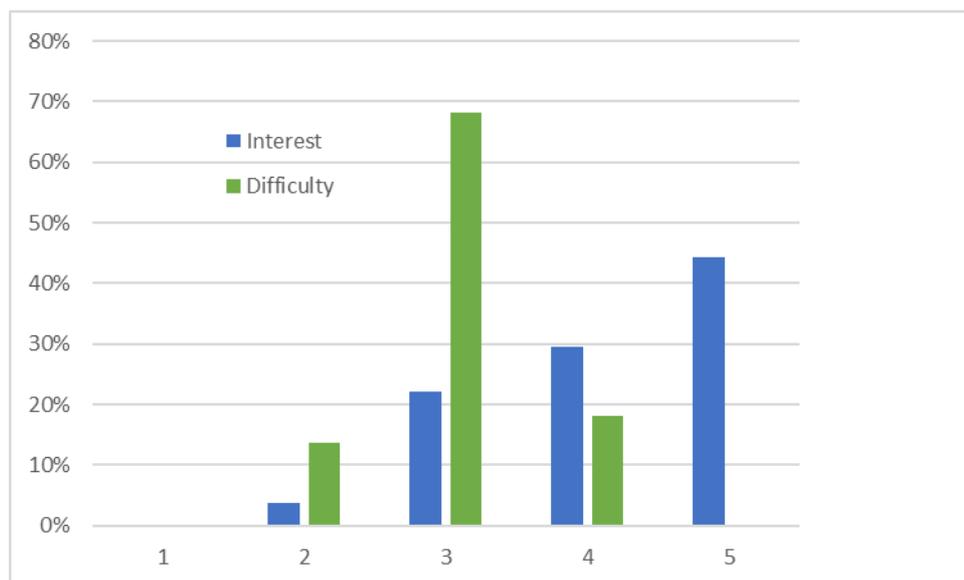

Fig. 1: How teachers rated their interest in CPS and the difficulty level of the activity proposed to the students (rank 1-5)

**Teachers' evaluation**

We asked the teachers to answer four questions: 1) did you find the CPS approach interesting? 2) what do you think about the distinction in roles? 3) what do you think about the distinction in steps? 4) what do you think about text-enriched problems?

In general, the evaluation was positive, almost the whole teachers' group rated the approach from quite to very interesting (Fig. 1). The same also holds for the other questions, being the different aspects of CPS rated quite or very efficient by most of the teachers (Fig. 2).

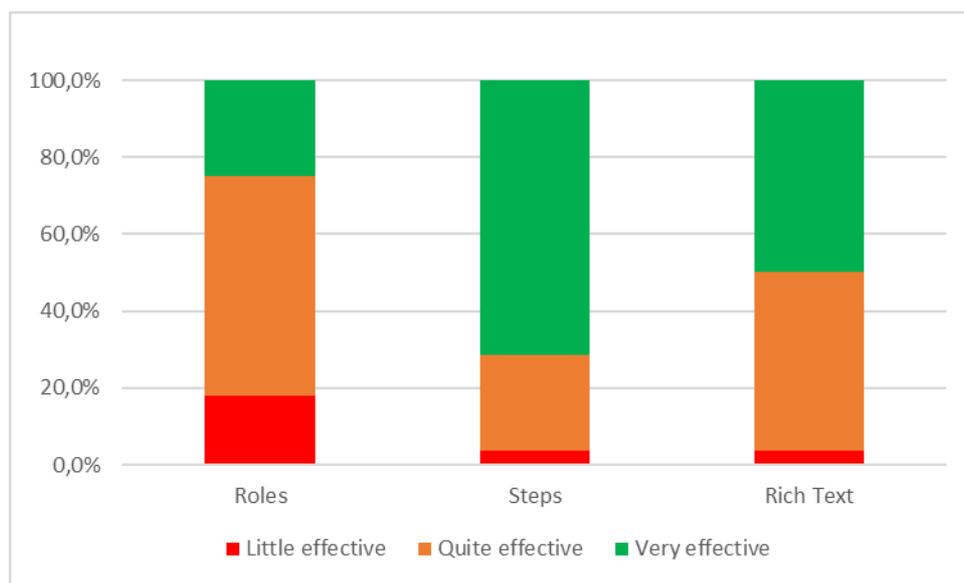

Fig. 2: How teachers evaluated different elements of CPS

As explained before, beside following lessons on the CPS method teachers were also involved in direct CPS experience divided in 3-member groups to evaluate the different aspects of the distinction in roles within the group and how to complete each separate step of an enriched problem. Finally, they were also engaged in the production and examination of enriched problems of physics, achieving at the end a common and shared database of problems. This experience allowed them to better understand the method and to evaluate its reliability in a class contest, suggesting an easier implementation in the final classes of the high school where the age and expected ripeness of students could make easier management and supervision of student groups. Indeed, in the experience with the students where they acted as tutor or coach, the difficulty level of the presented problems was rated as medium (Fig. 1) and the involvement of the students was evaluated as good (Fig.3). The teachers also evidenced a good capability of the students to understand the problem and to apply the solving method (the list of questions and grades is reported in table 1). It should be noted that teachers declared a minimal coaching activity, mainly devoted to simplifying understanding and separating the different steps of the method.

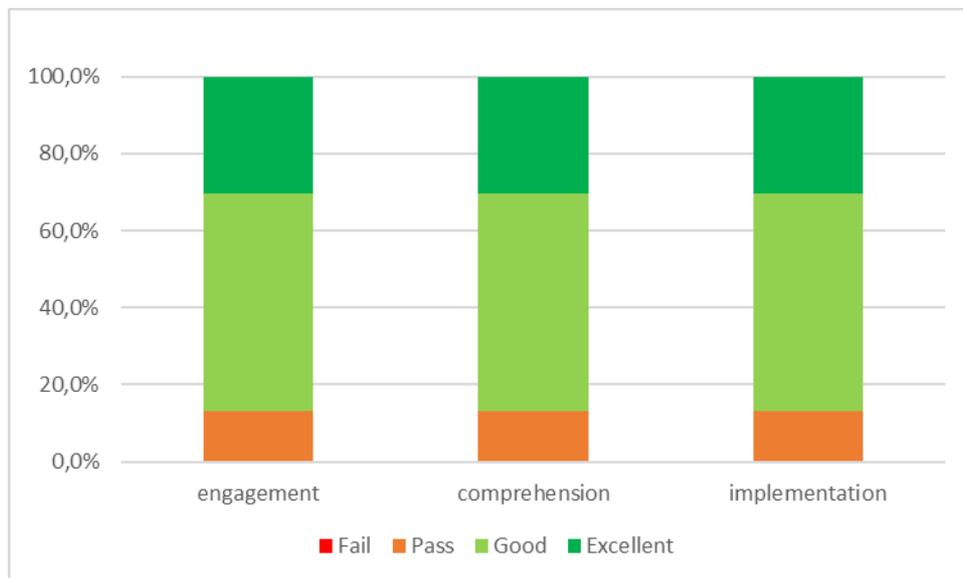

Fig. 3: Teachers evaluation of CPS experience with students

Table 1: List of questions posed to teachers (T) and students (S) to evaluate the CPS activity.

| Question | Ranking scale | Results |
|---|---|---|
|  |  |  |
| Difficulty level (T) | 1-5 | 2 (13.6%), 3 (68.2%), 4(18.2%) |
| Engagement (T) | 1-4 | 2 (13.1%), 3(56.5%), 4(30.4%) |
| Comprehension of the method (T) | 1-4 | 2 (34.8%), 3(34.8%), 4(30.4%) |
| Implementation of the method (T) | 1-4 | 2 (47.8%), 3(47.8%), 4(4.4%) |
|  |  |  |
| Comprehension of the problem (S) | Yes, partly or No | Yes (92.4%) partly (7.6%) |
| Comprehension of the step division (S) | Yes, partly or No | Yes (69.6%), partly (26.6%), No (3.8%) |
| Comprehension of roles (S) | Yes, partly or No | Yes (88.6%), partly (11,4%) |

| Preparatory School activity on CPS (S) | Yes, partly or No | Yes (29.1%) No (70.9%) |
|---|---|---|
| Evaluate step 1 (Focusing) (S) | 1 (unsolved)-3 (solved) | 2 (3.8%), 3 (96.2%) |
| Evaluate step 2 (Description) (S) | 1 (unsolved)-3 (solved) | 1 (1.2), 2 (32.5), 3 (66.3%) |
| Evaluate step 3 (Planning) (S) | 1 (unsolved)-3 (solved) | 2 (15%), 3 (85%) |
| Evaluate step 4 (Execution) (S) | 1 (unsolved)-3 (solved) | 1 (2.5%), 2 (23.8%), 3 (73.7) |
| Evaluate step 5 (Evaluation) (S) | 1 (unsolved)-3 (solved) | 1 (8.8%), 2 (31.2%), 3 (60%) |
| Evaluate your contribution as a single to solve the problem (S) | 1-5 | 1 (2.5%), 2 (5%), 3(28.8%), 4(51.2%), 5(5%) |
| Evaluate group work to solve the problem (S) | 1-5 | 1 (1.3%), 2 (3.7%), 3(30%), 4(38.8%), 5(26.2%) |
| Evaluate your contribution as a single to analyse the problem (S) | 1-5 | 1 (1.3%), 2 (3.7%), 3(31.3%), 4(47.5%), 5(16.2%) |
| Evaluate group work to analyse the problem (S) | 1-5 | 1 (1.3%), 2 (1.3%), 3(27.5%), 4(43.7%), 5(26.2%) |
| Evaluate your contribution as a single to plan the solution (S) | 1-5 | 1 (1.3%), 2 (6.2%), 3(25%), 4(47.5%), 5(20%) |
| Evaluate group work to plan the solution (S) | 1-5 | 1 (2.5%), 2 (5%), 3(21.3%), 4(36.2%), 5(35%) |
| Evaluate your contribution as a single to find/solve the proper equations (S) | 1-5 | 1 (1.3%), 2 (5%), 3(30%), 4(40%), 5(23.7%) |
| Evaluate group work to find/solve the proper equations (S) | 1-5 | 1 (2.5%), 2 (1.3%), 3(27.5%), 4(35%), 5(33.7%) |

**Students evaluation**

The evaluation was accomplished through a set of questions aimed to appraise the difficulty level, their comprehension of the method and level of cooperation. Finally, they were asked to self-judge their contribution and success.

In general, the students feel confident with their comprehension of the problem (Fig. 4-5) and their capability to apply the solution method, even though the matter was not prepared in class before attending the experiment (at least for the largest student group). Students also felt positive when appraising their contribution as a single or within the cooperation (both rated at least discrete in general, see supplementary) and estimated as successful their results in each step of the method, being their physics knowledge evaluated as suitable for the proposed problem (Fig. 4-5).

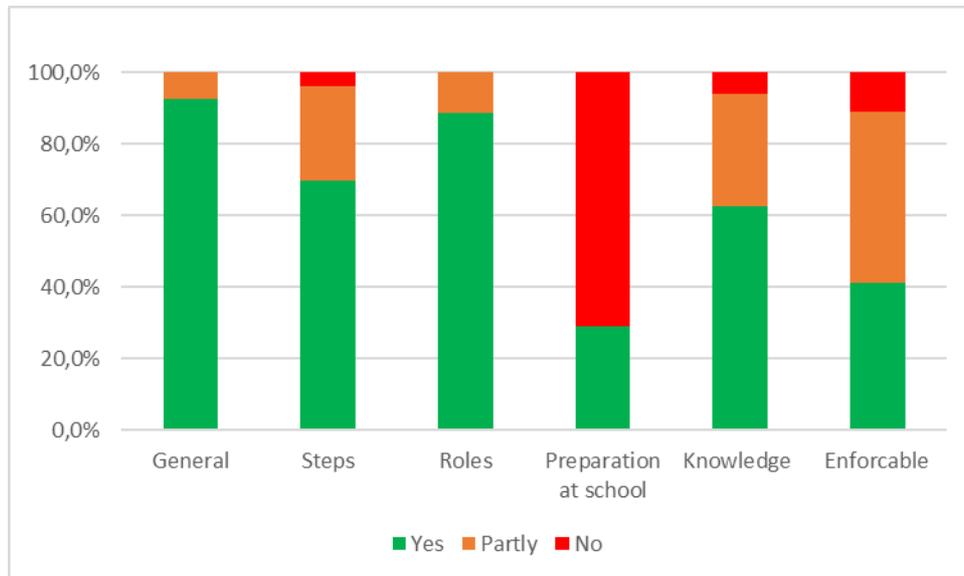

Fig. 4: Students evaluation of CPS experience

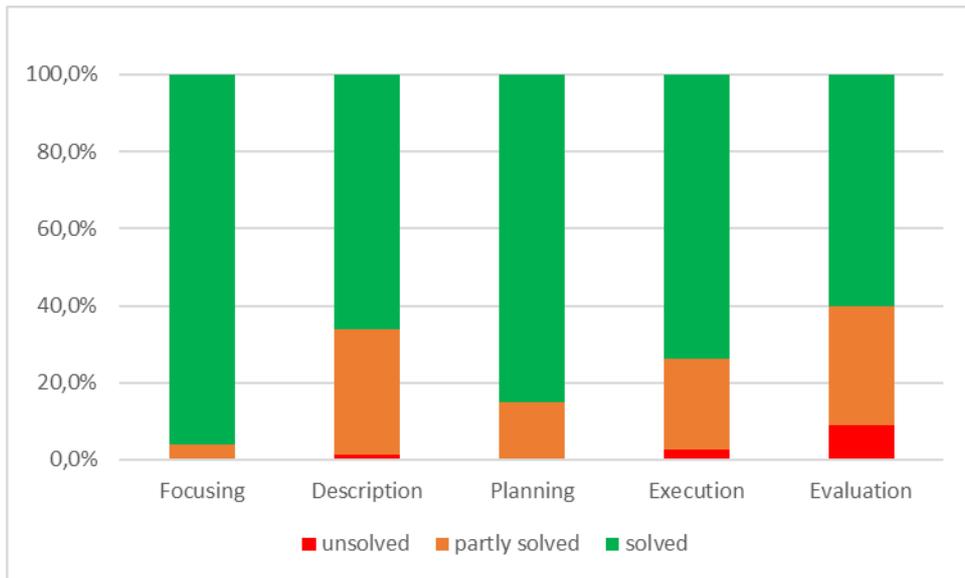

Fig. 5: Students self-evaluation of their success in different CPS steps

We also asked the students to express their comments on the experience, evidencing which were the aspects helping most to find the solution. The most appreciated aspect was the discussion within the group and the sharing of knowledge. Other beneficial aspects were the splitting in roles and the resolving scheme.

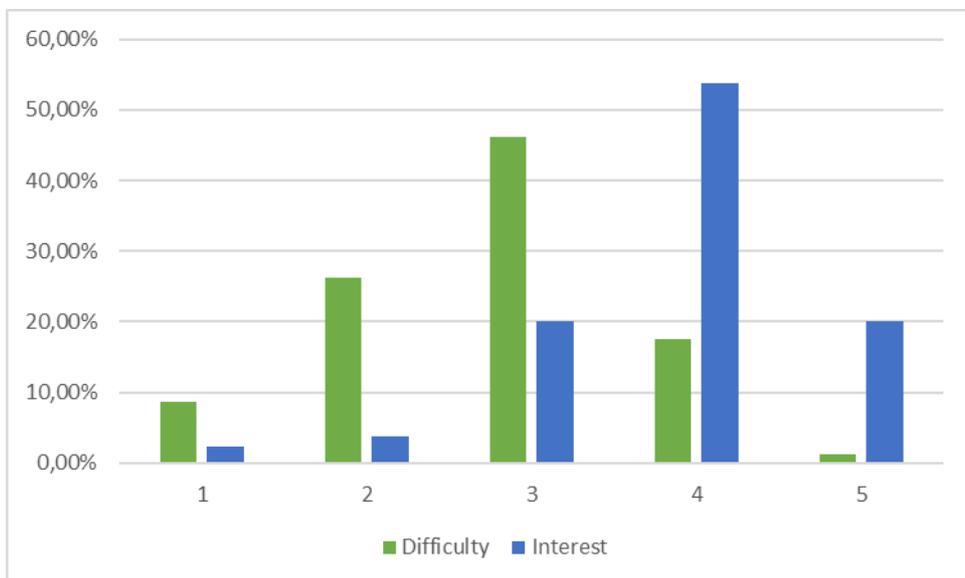

Fig. 6: Fig. 1: How students rated their interest in CPS and the difficulty level of the activity proposed to the students (rank 1-5)

Finally coaching was evaluated positively to ignite the discussion. However, the division in steps was found complex and somehow artificial, being not perfectly clear the separation among too much steps perceived as redundant. In general, the experience was evaluated as interesting (93.8% rated it >3 in a 1-5 scale) and with a medium level of difficulty (72.6% rated a level 2-3 in a 1-5 scale).

**Analysis of the solutions**

We analysed the elaborates of the students by ranking each step of the implemented solution scheme in a 0-1 rank. The results are collected in the graph by grouping as insufficient, sufficient and good the 0.0-0.4, 0.5-0.7 and 0.8-1.0 ranges.

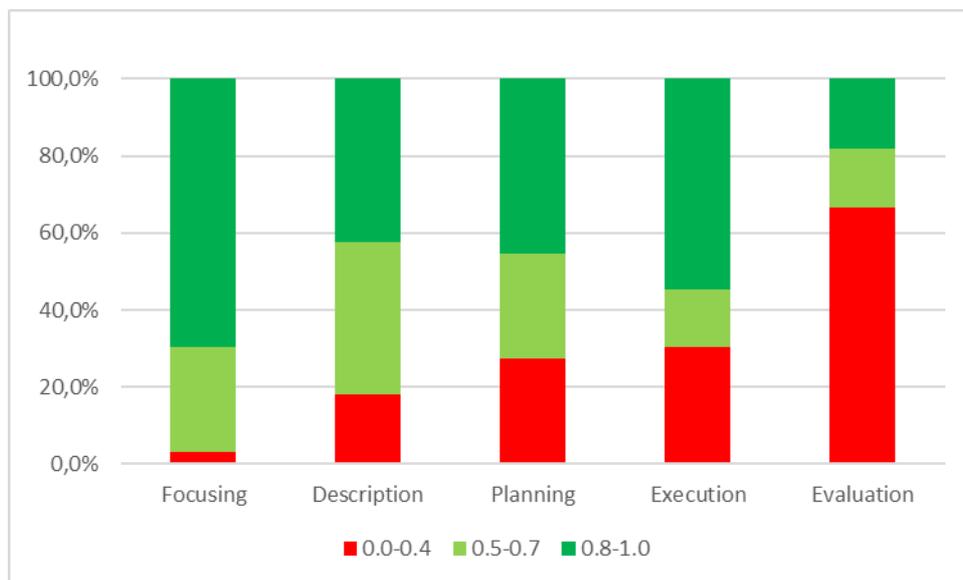

Fig. 7: Analysis of the students elaborates

As reported in Fig. 7 students experimented increasing difficulties in the different steps of the solution scheme but more than 80% of them were able to successfully complete the first two steps (focusing and description) and up to 70% the plan and execution steps. The percentages are totally reversed in the last step (evaluation) where the students should evaluate their results and give reasons if they found them reasonable or not. The most part of the groups did not understand the request of evaluation and at most gave the easy answer as yes or no, despite they were advised that any results should be examined in a rational basis. It should be noted that this is in contrast with the student feeling of successfully reaching the solution. It indicates, in our opinion, that the general approach of the students to solve a problem is to find a number, with no further speculation on the reliability and soundness of the found number, evidencing a general lack of the capacity of abstraction and generalization. This was already reported in previous studies[4] and in general refers to the different approach of expert and novice to problem solving[12, 13, 22]. Improving their critical sense is a crucial aspect to increase their problem-solving ability, allowing conversion of novices into experts and helping the students in developing a more objective self-analysis of their performances.

This analysis is also confirmed when one takes a look at the gender mean vote distribution (see Fig. 8) in students during the course.

Although the cooperative problem-solving course we proposed was not focused on the analysis of results of single students in a group, it is interesting to evaluate differences (if there are some) in gender performances obtained by students. To do so, we considered groups' vote in every CPS step and then, we decided to equally distribute them among the participants to the group. This choice is certainly not unique; however, it is the most conservative and reliable one for our purposes. Our sample is composed by 34 female students and by 64 male students.

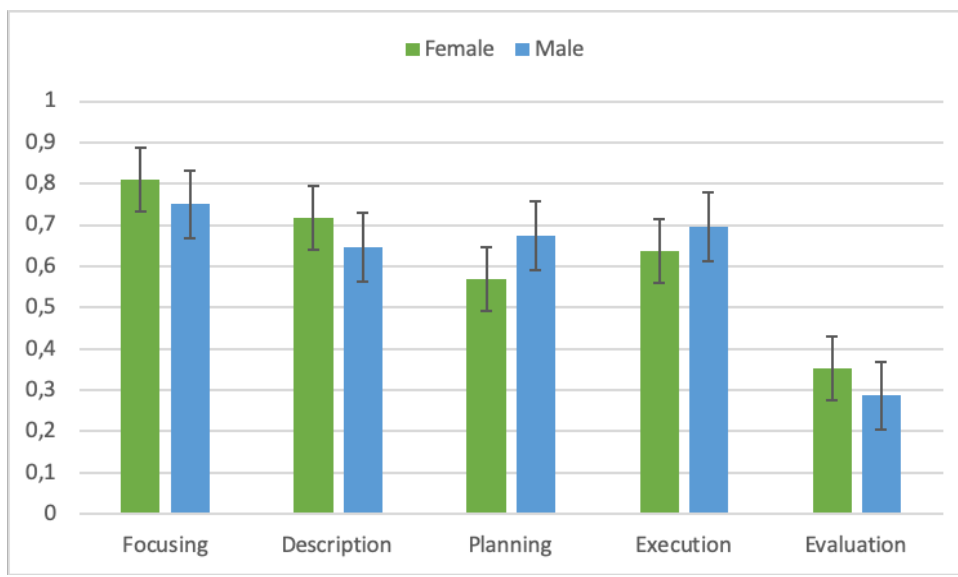

Fig 8: Mean Vote Distribution per Gender. Error bars represent the statistical error in mean vote distribution data.

The results of the analysis are shown in Fig. 8. Female students obtained better results than male students in focusing and describing the problem they are looking at. Male students were more efficient than females' colleagues in planning strategies and executing calculation to find solutions to problems. For what concerns the last step, evaluation, the results are comparable for the twos. However, in females it emerges a slight increased capability to evaluate the goodness of the obtained results with respect to male students. The results of our analysis confirm what is well known in literature about gender differences in physics problem solving[6, 25, 26].

The last point of our analysis focuses in comparing results obtained by students coming from scientific and classical high schools (Fig. 9). In order to do so, as in the previous case, we examined results of single students by equally distributing group's votes in every CPS step among single students of the group. Then, we collect votes in two different categories, the ones related to students coming from scientific high schools and the ones from classical high schools.

Let us remark that our data sample is composed by 98 students, the large part (84%) coming from scientific high schools (8 schools), whereas the others (16%) come from classical high schools (1 school).

The results of the analysis are shown in Fig. 9.

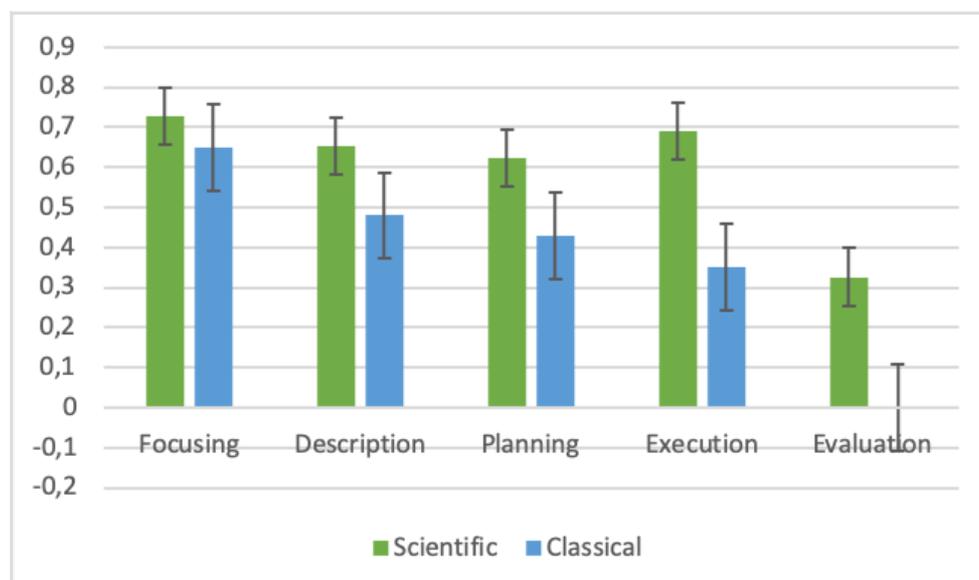

Fig. 9: Mean vote distribution among scientific and classical high schools' students. Error bars represent the statistical error in data.

What emerges from our analysis is that students coming from scientific high school obtained better results with respect to their classical colleagues. This can be related to the intrinsic differences in the institutional programs in physics between the two types of institutes and, therefore, in the way professors teach physics.

Scientific high schools are, by definition, devoted to science and students are more familiar with respect to their colleagues coming from classical institutes to solving exercises like the ones we proposed during the course. For these reasons, as expected, the average capability of facing up with problem solving is statistically more prominent in scientific high school students than classical ones.

We can also observe that in the case of students coming from scientific schools, the absolute distribution of votes has the same trend of the one shown in Fig. 7. This can, in principle, reflect the fact that the majority of the students in our sample comes from scientific institutes. Nevertheless, quite interesting, for classical high schools' students the distribution of votes shows a negative trend. Moreover, the gap between their votes and the ones reported by their scientific colleagues in the last two CPS steps increase. Presently, we do not know if this result reflects a problematic associated in teaching physics in classical high schools or if it is just related to a (rather unlucky) statistical distribution (of votes) of students at the course. However, this is an interesting result we leave for future investigations.

Finally, we want to draw our attention on the "evaluation" step in Fig. 8. In the case of scientific high schools, the result confirms what is emerged from the analysis of CPS working groups: students have shown difficulties in evaluating the results they have found in solving physics problems. Conversely, when one takes a look at the mean vote distribution per type of institute, immediately realize that students coming from classical high schools have obtained zero in this step, meaning they are not able to evaluate their results at all. Of course, also in this case, an explanation like the one reported above could be plausible, but it seems at least too simplistic to be valid. This point deserves a more accurate analysis with a dedicated investigation we leave for future studies.

**Conclusions**

We proposed the cooperative problem-solving technique (CPS) to high-school physics teacher and discussed its applicability to Italian high-school classes by performing an experience of CPS where the teachers acted as solvers. We also simulated an application of the method to final classes students to verify how the students evaluate the new technique (in this case the teachers acted as tutors or coaches). Teachers appreciated the method and suggested that final classes could be the proper ones where the method could be introduced because of the need of abstraction and speculation. The most appreciated aspects were the group working and text rich context, evaluated as really positive in stimulating student engagement, even though preparation of rich context problems requires lot of effort. The students appreciated the same aspects but perceived the problem division in different steps and somehow the role splitting as a compelling over structure. The analysis of their performance displayed a quite good success, considering that there was not, in general, previous preparation and it was their first attempt in CPS. However, the analysis displayed also some difficulties in separating the different steps of the methods, despite the use of a solution scheme, and, above all, showed a large fault in the self-evaluation process and in the evaluation of the reached results. These findings show that there is a large need to develop critical sense and abstraction ability of students to improve their problem solving skills, results which could be achieved by CPS implementation in high-school classes.

Finally, we have examined the gender and school's distribution of votes in students during the CPS course. In the former case, we have found a confirmation of previous analysis about gender differences in physics problem solving[6, 25, 26]. The analysis have shown that females students are more capable than male ones in focusing and describing the problem they are looking at. On the contrary, male students have been more efficient than females' colleagues in planning strategies and executing calculation to find solutions to problems. For what concerns the CPS step called "evaluation", the results are comparable for the twos. However, a slight increased capability to evaluate the goodness of the obtained results appears in females' data with respect to males' ones. For what concerns the schools' distribution of votes, the sample was composed by students coming from scientific (84%) and classical (16%) high schools. In general, students coming from scientific high schools have reported better results than their classical colleagues. In general, we expected this result due to the different intrinsic nature of the two institutes and to the different pedagogical approaches they have in didactic of science. However, differently to their scientific colleagues, the vote distribution in classical high schools' students show a negative trend. In particular, it emerges a complete incapability in evaluating the results they have obtained in solving physical problems. These two points would deserve a separate analysis we leave for a future work.

Supplementary File

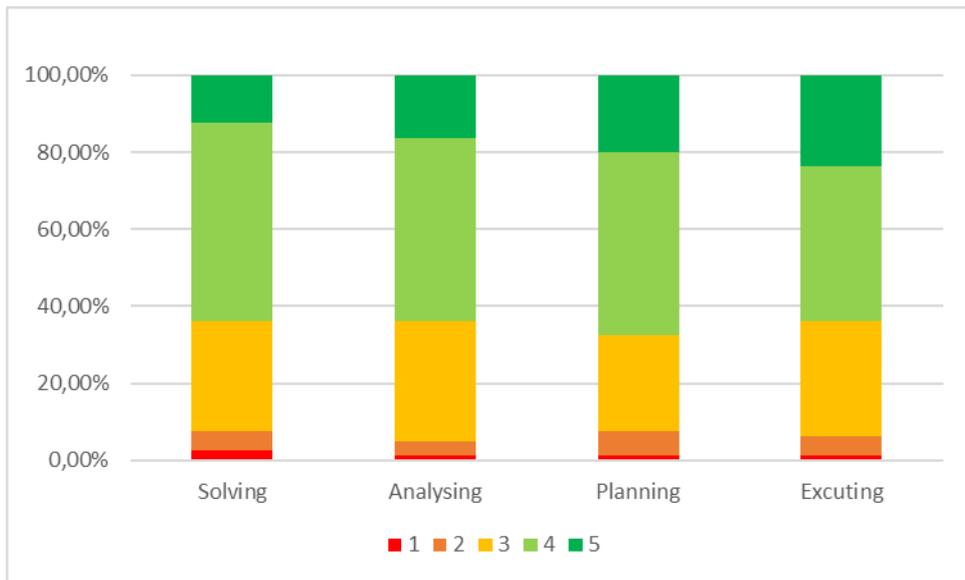

Fig. S1: Students evaluation of their contribution as individual to problem solution

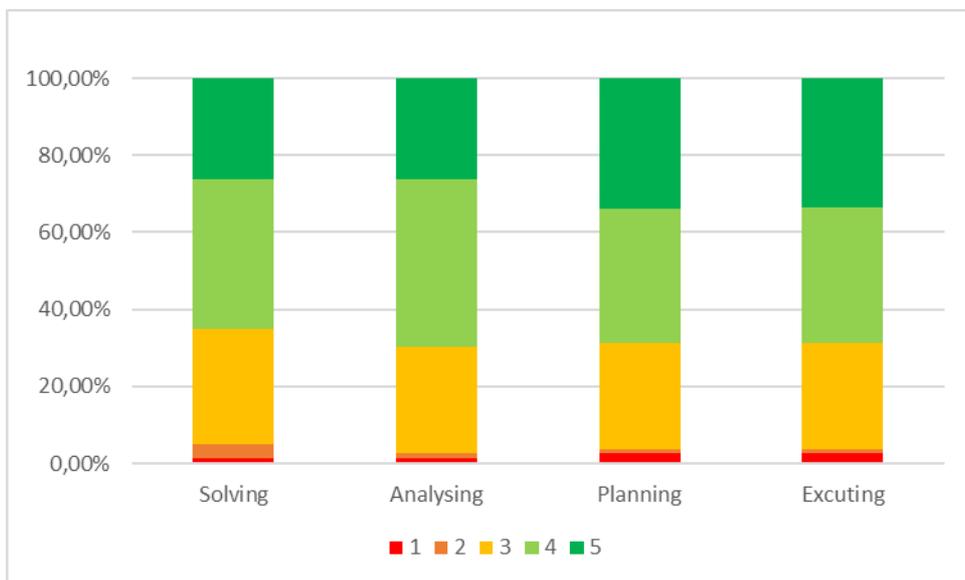

Fig. S2: Students evaluation of their contribution as group to problem solution